# Overlay journals: A study of the current landscape

Antti Mikael Rousi
Aalto University, Finland

Mikael Laakso
Hanken School of Economics, Finland

## Abstract
Overlay journals are characterised by their articles being published on open access repositories, often already starting in their initial preprint form as a prerequisite for submission to the journal prior to initiating the peer-review process. In this study we aimed to identify currently active overlay journals and examine their characteristics. We utilised an explorative web search and contacted key service providers for additional information. The final sample consisted of 34 overlay journals. While the results show that new overlay journals have been actively established within recent years, the current presence of overlay journals remains diminutive compared to the overall number of open access journals. Most overlay journals publish articles in natural sciences, mathematics or computer sciences, and are commonly published by groups of academics rather than formal organisations. They may also rank highly within the traditional journal citation metrics. None of the investigated journals required fees from authors, which is likely related to the cost-effective aspects of the overlay publishing model. Both the growth in adoption of open access preprint repositories and researchers' willingness to publish in overlay journals will determine the model's wider impact on scholarly publishing.

## Introduction

Digitisation of the scholarly publishing industry in the 1990s enabled major efficiency gains compared to the operating circumstances dictated by the paper-based past. Large publishers were able to benefit from even greater economies of scale, both in production and sale of read-access to published materials; for smaller actors, the cost and technical competency required to run a journal have also become lower with the maturation of open source solutions like Open Journals Systems. However, the functional logic and service offering of journals have remained largely the same despite the shift to digital. So far, growth in open access (OA) through journals has largely been driven by outlets that adhere to manuscript handling, peer-review and publication processes that are not radically different from print-based processes. Alternative publishing models that more fundamentally leverage the circumstances of digital-only and OA have been evolving. Among them is the overlay journal model (Herman et al., 2020) – the core focus of this study. Although the definition of what constitutes an overlay journal may vary (Brown, 2010), overlaying peer-review and journal structures on top of already existing preprint repository infrastructures are usually at the core. This study considers the following criteria for an overlay journal: for peer-reviewed original articles, the journal (1) mandates or encourages the submission of preprints to an external OA repository and (2) hosts the final published articles in an external OA repository.

The overlay journal is an emerging model within academic journal publishing, but there is nothing conceptually or practically new. In terms of widespread uptake, it has already been forming – albeit obscurely – for the last two

**Corresponding author:**
Antti Mikael Rousi, Research Services, Aalto University, Otaniementie 9, Aalto 00076, Finland.
Email: antti.m.rousi@aalto.fi

decades (Pinfield, 2009; Smith, 2000; Thornton and Kroeker, 2021). Earlier literature has referred to such journals as deconstructed journals or superjournals (Eysenbach, 2019; Smith, 1999). Overlay journals combine elements from the gold and green routes to OA, which have often been presented as separate paths to enable OA. On one hand, the gold route is commonly depicted as publisher-provided OA, where the final published articles are freely available through the journal's website to any interested reader (Laakso and Björk, 2013). On the other hand, the green route is author-provided OA, where the manuscript version of the work is deposited into an OA repository or to the researcher's personal website from which it is freely available to any interested reader (Laakso and Björk, 2013). With the overlay model, the gold OA element is reflected in the journal curating content through editorial work, managing its peer-review and ultimately making the final output available OA without any paywall for readers. The green element comes from overlay journals basing parts of their publication processes on public OA repositories (see e.g. Pinfield, 2009). As reviewed in closer detail as part of this study, the processes of manuscript submission and peer-review management differ among overlay journals. However, in the most common form, authors first upload their preprint to an OA repository and the link to the manuscript is sent to the journal editors. After peer-review and formal acceptance, which is managed by the journal editors, the authors upload the final article version to the OA repository. The overlay journal links to this and provides journal issue and volume information (hence 'overlay').

It can be argued that the growth of preprints and the recent increased momentum around overlay journals are interconnected. Similar to overlay journals, the practice of posting preprint versions of journal article manuscripts is not new. It has been around for several decades – even before the world wide web – but the COVID-19 pandemic has fuelled new growth and acceleration of the discussion on this topic that concerns a broad spectrum of research disciplines (Fraser et al., 2021). The growth of preprint distribution has been challenged by compatibility with the journal publishing sector, where many journals have unclear policies on permissibility or restrict such practices for manuscripts submitted to the journal (Klebel et al., 2020). Studies on the connection between preprints, journal submission and publication reveal diversity in how and when authors upload preprints, with most uploads occurring around the time that authors submit their manuscript to a journal (Anderson, 2020; Larivière et al., 2014).

The overlay journal model can potentially enable several benefits over traditional ways of publishing and communicating research results. Overlay journals are highly compatible with and build upon the growing practice of researchers making preprints of their manuscripts available prior to formal peer-review of the content, thus potentially bringing more speed and openness to the research communication processes compared to publication of formal journal outputs. The use of existing digital repositories and their technology has been seen to reduce publishing costs related to, for example, content acquisition, dissemination and archiving for overlay journals (Grossmann and Brembs, 2021: 3). The overlay model may thus be seen as a cost-effective model of OA publishing that 'piggy-backs' on the technology already provided by digital repositories (Grossmann and Brembs, 2021: 10). For self-managed journals, the issue of preservation of published materials does not happen automatically; indeed, some journals have disappeared together with their published materials (Laakso et al., 2021). Preservation of published materials has the potential to be well managed through the overlay publication model, as long as the utilised repositories perform robust backups and plan for long-term archival of materials. As there is yet a lack of widely adopted standards and practices, it is important that repositories transparently communicate aspects related to their sustainability for ease of assessment of their trustworthiness (Lin et al., 2020).

The key research articles looking into overlay journals are already over a decade old (Pinfield, 2009; Smith, 2000), which is a long time considering the pace of change in this space. An exception to this is a recent conference paper by Thornton and Kroeker (2021), where the authors summarise the state and growth of the largest subject repositories and compiled a list of past and presently active overlay journals together with their key characteristics. The present work extends upon this foundation by focussing on currently active overlay journals, analysing their characteristics in depth, including manuscript source repositories, potential article processing charge for authors, research disciplines, publisher types, content licencing, publication volume, material submission and peer-review processes, as well as key indexation and citation metrics.

The following terminology is at the core of this research. *Preprint repositories* refer to OA digital repositories that host and disseminate various versions of journal articles. The preprint repositories (sometimes also referred to as preprint servers or e-prints) used by the journals examined in the present study include repositories such as arXiv and HAL. *Overlay journal service provider* refers to the provider of the web service that enables linking and showcasing the documents hosted in preprint repositories for compiling journal issues, and provides information for the article including issue, volume and DOI. Although the definition of an *overlay journal* varies (Brown, 2010), this study utilises the following criteria: for peer-reviewed original articles, the journal (1) mandates or encourages the submission of preprints to an external OA repository and (2) hosts the final published articles in an external OA repository. Given that overlay journals may be published by groups of academics rather than formal organisations, the present study approached the concept of *publisher* in a broad sense. The investigated overlay journals are scrutinised from the viewpoint of the nature of the organisation behind publications, that is, whether they are published by formal commercial or non-commercial organisations, or by more informal communities such as groups of academics.

## Methodology

### Research aim

The research aim of this article is to identify all journals currently operating with an overlay model, and to examine their characteristics. The main research questions are posed as follows:

- RQ1: What journals are currently operating with the overlay model, and what are their fields of research?

Once the current overlay journals are identified, the present study seeks to analyse their characteristics as follows:

- RQ2.1: When were the journals established, and when did they start publishing based on an overlay model?
- RQ2.2: Are there article processing charges for authors publishing in the journals?
- RQ2.3: Do the journals practice open peer-review?
- RQ2.4: What are the policies for submissions and preprints?
- RQ2.5: Which preprint repositories are journals using as part of their overlay model?
- RQ2.6: What are the five largest overlay journals by article volume, and how has their output developed over the last 4 years?
- RQ2.7: What are the publication languages?
- RQ2.8: What are the publisher types?
- RQ2.9: What overlay journal service providers are used?
- RQ2.10: What licences are used for published content?
- RQ2.11: How well are the journals indexed in DOAJ, Scopus and WoS?
- RQ2.12: What do the journal metrics look like?

### Data collection

Since no definitive list of overlay journals exists, identifying the sample journals was an explorative process (occurring during April 2021–February 2022). The sample of investigated overlay journals were identified by using the websites of Episciences.org (2021), Scholastica (2021), Free Journal Network (2021), Open Journals (2021), PubPub (2022) and Wikipedia (2021). The Scholastica and PubPub teams kindly provided information about their overlay journals to the authors. Twitter (2021) was searched with terms such as 'arxiv overlay' and 'overlay journal' to identify additional journals. In total, this study identified 34 overlay journals currently accepting submissions.

Platform-based venues hosting preprints and final articles on their own platform, such as *F1000 Research, PeerJ, Scipost* and *Open Research Europe*, were not included in this study. In addition, journals publishing software that refer to GitHub for source code, such as *Journal of Open Source Software*, were not treated as overlay journals in the present study. Furthermore, rapid review and annotation type of venues, such as *Rapid Reviews Covid-19* were excluded from outlets included in this study. Although these venues utilise OA repositories and facilitate peer-review of preprints, they do not incorporate a structured peer-review process that would lead to an authoritative editorial decision for inclusion or exclusion of materials in the venue. In addition, journals that previously operated using the overlay model, but are no longer active, that is, do not accept manuscript submissions (e.g. *BiOverlay*), or have transferred to different publishing models (e.g. *Journal of High Energy Physics*), were excluded from the study. There are further two initiatives that fulfil some elements of our definition for overlay journals but not all and are thus not included. These services are discussed in the two following paragraphs.

The *Peer Community In* (PCI) is an initiative currently experimenting with combining repository functions with publication of scientific journals. Thus far, the PCI has set up 15 thematic groups that are open to submissions of preprints from any public repository for an open peer review on authors' request. Once peer-review is completed and revisions are made, thematic group members can formulate a public recommendation for the preprint. While PCI-recommended articles can continue to reside in a repository, the initiative also offers other paths. For example, the *Peer Community Journal*, established in 2021, accepts all PCI-recommended articles. The *Peer Community Journal* is distinct from the overlay journals examined in the present study because the final article versions are not necessarily hosted on an external repository. This is due to *Peer Community Journal* accepting preprints from any public repository, some of which may not allow adding information about the work being published in a journal, for example. Alternatively, authors can submit their PCI-recommended article to 1 of the 90 journals in various disciplines that have signed up for accepting PCI-recommended articles, or have agreed to provide a rapid response if other editorial decisions are made. Another option is to submit the PCI-recommended article to any other journal (Peer Community in Initiative, 2022).

*Social Science Research Network* (SSRN) *eJournals* is also an important initiative experimenting with combining preprint repository functions with traditional journal publishing. *SRRN First look journals* provide the opportunity for authors of selected journals to publish their preprint in SRRN while the manuscript is under peer-review. However, as the final articles are not hosted OA in SRRN, these journals were not considered as overlay journals in the present study. To the authors' best knowledge, the remaining *SSRN eJournals* either require fees for accessing the final article versions (either for all or selected articles) or are OA journals that use SSRN for content dissemination without a defined preprint policy. For these reasons, *SSRN eJournals* were not examined as overlay journals in the present study.

## Data analysis

### Journal characteristics

The journal ISSN numbers, manuscript source repositories, first overlay volumes, article volumes, publication languages, peer-review type, licence for published articles, author costs, publisher types, submission policy and preprint availability policy were observed by inspecting journal editorial policies and submission guidelines found from journal websites. The overlay journals' ISSN numbers were identified by examining journal websites and cross-checking this information with the Ulrich's periodicals database (Ulrichsweb, 2021). Journals that published review reports, either with reviewers' names or anonymously, were classified as operating with open peer-review. Publisher types defined by Laakso and Björk (2013) were used to categorise the findings concerning the publishers. If the journal website did not include publisher information, the editorial board was interpreted to publish the journal.

As the sample journals were not comprehensively covered by any catalogue providing field of science information (e.g. Ulrichsweb database or DOAJ), the Organisation for Economic Co-operation and Development (OECD) (2007) field of science classification was used to categorise the journals into different domains of science. The journals' primary OECD field of sciences were defined by the authors through examining the journal websites. Exceptionally, the *ST-Open* and *Journal of Interdisciplinary Methodologies and Issues in Science* journals were given a multidisciplinary classification, which as a category is not listed in the OECD field of science classification.

### Journal indexing

Whether the journals were indexed in the Directory of Open Access Journals (DOAJ), Scopus or Clarivate Analytics' Web of Science Core collection's journal master list was examined by searching the services with journal ISSN numbers and journal titles (Clarivate, 2021a; Directory of Open Access Journals [DOAJ], 2021; Elsevier, 2021). If a journal was found to be indexed in the Scopus database, the article volumes for the years of 2018–2021 were extracted. For journals and volumes not indexed in Scopus, articles were manually counted from the journal webpages.

### Journal metrics

The identified overlay journals were examined from the viewpoint of both qualitative and quantitative journal metrics. The qualitative metrics comprised the Nordic expert panel rankings of scientific journals, namely the Finnish Publication Forum (FPF) (Finnish Federation of Learned Societies, 2018), the Danish Bibliometric Research Indicator (DBRI) (Ministry of Higher Education and Science, 2022) and the Norwegian Register for Scientific Journals, Series and Publishers (NRSJP) (Norwegian Directorate for Higher Education and Skills, 2022). The Finnish Publication Forum and the Danish Bibliometric Research Indicator classifications place journals into one of three levels: 1 = basic; 2 = leading and 3 = top. The Norwegian Register for Scientific Journals, Series and Publisher operates with two levels of classification: 1 = basic and 2 = leading. It is noteworthy that these classifications include journals by suggestions from the public, so exclusion from the listed journals does not mean that the journals could not be considered as scientific. Searches were conducted from the web portals of the above services with both ISSN numbers and journal titles.

To add a quantitative citation metric into the investigation, Clarivate Analytics' Journal Citation Reports database (Clarivate, 2021b) was searched with the use of both ISSN numbers and journal titles to identify whether the journals had a Journal Citation Indicator (JCI), 2-year Impact Factor (IF) and an Impact Factor ranking (IF rank). The examined Journal Citation Indicators, Journal Impact Factors and Impact Factor rankings were for the year 2020 (as released in 2021). The collected data are provided as open data to facilitate future research.

The data are openly archived here https://doi.org/10.5281/zenodo.6420517.

## Findings

### RQ1: What journals are currently operating with the overlay model, and what are their fields of research?

This study identified 34 overlay journals currently accepting submissions, of which 29 (85%) had an ISSN number. Table 1 presents a listing of the identified journals categorised by research field.

**Table 1.** Journal titles, ISSN numbers and URLs sorted into research field categories.

| Journal name | ISSN | Website |
|---|---|---|
| *Basic Medicine* | | |
| Neurons, Behaviour, Data Analysis and Theory | 2690-2664 | https://nbdt.scholasticahq.com/ |
| Mathematical Neuroscience and Applications | 2801-0159 | https://mna.episciences.org/ |
| *Biological Sciences* | | |
| JMIRx | Bio | Not found | https://bio.jmirx.org/ |
| *Clinical Medicine* | | |
| JMIRx | Med | 2563-6316 | https://med.jmirx.org/ |
| *Computer and Information Sciences* | | |
| Machine Learning for Biomedical Imaging | 2766-905X | https://www.melba-journal.org/ |
| African Journal of Research in Computer Science and Applied Mathematics | 1638-5713 | https://arima.episciences.org/page/politiques-editoriales |
| Discrete Mathematics and Theoretical Computer Science | 1365-8050 | https://dmtcs.episciences.org/ |
| Fundamenta Informaticae | 1875-8681 | https://fi.episciences.org/ |
| Journal of Data Mining and Digital Humanities | 2416-5999 | https://jdmdh.episciences.org/ |
| Journal of Interaction Between Persons and Systems | 2418-1838 | https://jips.episciences.org/ |
| Logical Methods in Computer Science | 1860-5974 | https://lmcs.episciences.org/page/authors-information |
| The Art, Science and Engineering of Programming | 2473-7321 | https://programming-journal.org/ |
| TheoretiCS | Not found | https://theoretics.episciences.org/ |
| *Economics and Business* | | |
| Journal of Philosophical Economics | 1844-8208 | https://gcc.episciences.org/ |
| Management et organisations du sport | 2804-8598 | https://mos.episciences.org/ |
| *Languages and Literature* | | |
| Slovo | 2557-9851 | https://slovo.episciences.org/ |
| *Mathematics* | | |
| Advances in Combinatorics | 2517-5599 | https://www.advancesincombinatorics.com/ |
| Internet Mathematics | 1944-9488 | https://www.internetmathematicsjournal.com/ |
| Discrete Analysis | 2397-3129 | https://discreteanalysisjournal.com/ |
| Ars Inveniendi Analytica | 2769-8505 | https://ars-inveniendi-analytica.com/ |
| Épijournal de Didactique et Epistémologie des Mathématiques pour | Not found | https://epidemes.episciences.org/ |
| Épijournal de Géométrie Algébrique | 2491-6765 | https://epiga.episciences.org/ |
| Hardy Ramanujan Journal | Not found | https://hrj.episciences.org/ |
| Journal of Groups, Complexity, Cryptology | 1869-6104 | https://gcc.episciences.org/ |
| Journal of Nonsmooth Analysis and Optimisation | 2700-7448 | https://jnsao.episciences.org/ |
| *Mechanical Engineering* | | |
| Journal of Theoretical, Computational and Applied Mechanics | 2726-6141 | https://jtcam.episciences.org/ |
| *Multidisciplinary* | | |
| Journal of Interdisciplinary Methodologies and Issues in Science | 2430-3038 | https://jimis.episciences.org/ |
| ST-Open | 2718-3734 | http://st-open.unist.hr/index.php/st-open |
| *Philosophy, Ethics and Religion* | | |
| Sociétés plurielles | 2557-9959 | https://societes-plurielles.episciences.org/ |
| *Physical Sciences* | | |
| The Open Journal of Astrophysics | 2565-6120 | https://astro.theoj.org/ |
| The Open Journal for Quantum Science | 2521-327X | https://quantum-journal.org/ |
| Symmetry, Integrability and Geometry: Methods and Applications | 1815-0659 | https://www.emis.de/journals/SIGMA/ |
| Open Communications in Nonlinear Mathematical Physics | 2802-9356 | https://ocnmp.episciences.org/page/for-authors |
| *Psychology* | | |
| JMIRx | Psy | Not found | https://psy.jmirx.org/ |

The two research areas with the most journals were computer and information sciences (*n*=9) and mathematics (*n*=9). This was followed by physical sciences (*n*=4) and basic medicine (*n*=2), economics and business (*n*=2), multidisciplinary (*n*=2), biological sciences (*n*=1), clinical medicine (*n*=1), languages and literature (*n*=1), mechanical engineering (*n*=1), philosophy, ethics and religion (*n*=1) and psychology (*n*=1). Since journal sizes can vary considerably, we investigated the number of published articles in each journal during the 4-year period of 2018–2021 to gain a better understanding of the differences among fields of research (Figure 1). The four overlay journals in the physical sciences published the most articles in 2021 (364 in total) with an increasing trend over the last few years, while the eight journals in computer and information sciences showed a fairly stable article output over the 4 years (total of 166 in 2021). The eight mathematics journals provided a stable annual output of around 50 articles over the 4-year period.

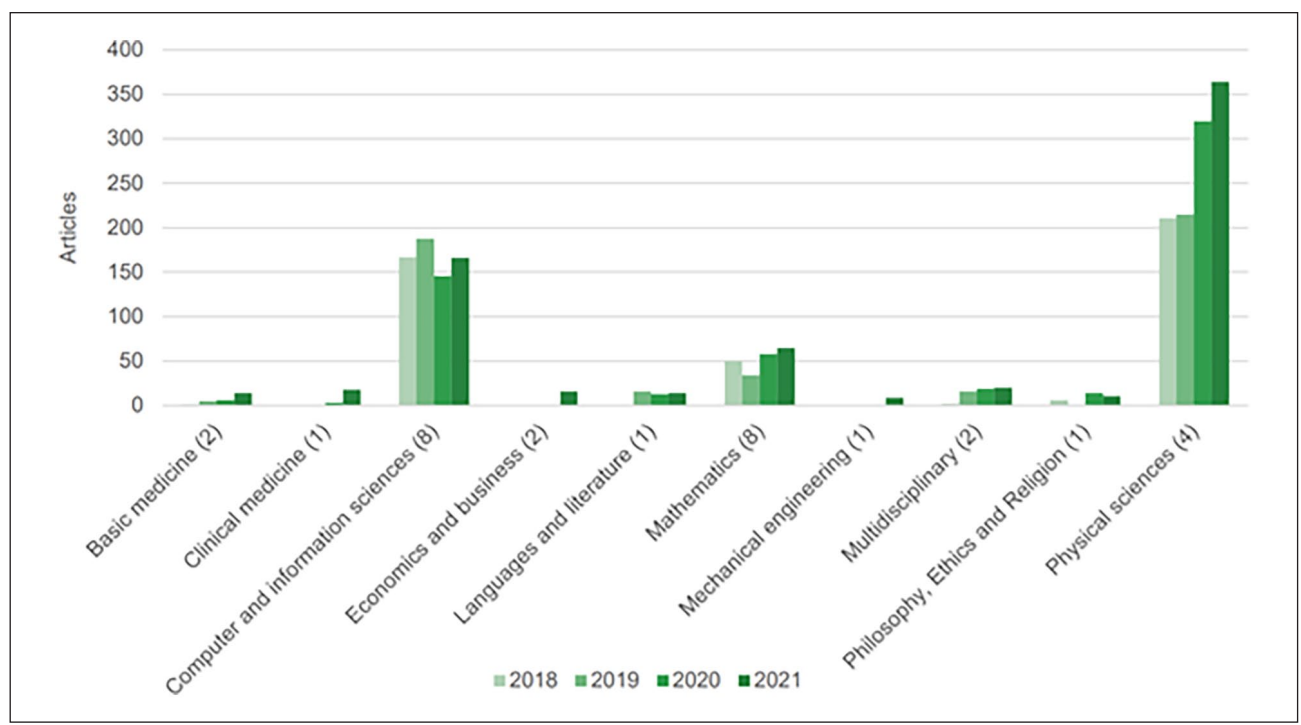

**Figure 1.** Total article output for overlay journals per research field category for the years 2018–2021. Only articles published based on an overlay model during each year are included in the totals. The number in brackets indicates how many journals contributed towards the total article count.

### RQ2.1: When were the journals established, and when did they start publishing based on an overlay model?

The previous research question alluded to the fact that overlay journals have not necessarily adopted the model from their inception, but rather have converted to it at some point in their lifecycle. Figure 2 visualises the year of journal establishment and when it was converted to an overlay model. Some journals such as the *Hardy Ramanujan Journal* have retrospectively uploaded articles to an OA repository after making the transfer to the overlay model. The results indicate a broad mix of backgrounds and research fields contributing to the current composition of journals in the landscape. It is notable that almost half of the journals ($n = 16$; 47%) started as overlay journals from 2020 or later.

### RQ2.2: Are there article processing charges for publishing in the journals?

Our analysis revealed that the investigated overlay journals did not require fees either from readers or authors. *The Open Journal for Quantum Science* had a voluntary article processing charge for the authors. Thirty (88%) of the investigated journals explicitly stated having a no-fee policy. Three (9%) of the journals did not provide information about author charges on their websites, which we assumed to mean that no payment would be required.

### RQ2.3: Do the journals practice open peer-review?

A clear majority of the investigated journals operated using peer-review that remained blind throughout the entire process ($n = 29$; 85%). Only 1 (3%) of the investigated journals reported to conduct open peer-review by publishing reviewer reports. In addition, 2 (6%) of the journals accepted both blind and open peer-reviews, depending on the preference of authors. Two (6%) of the investigated journals did not explicitly define their peer-review process or policy.

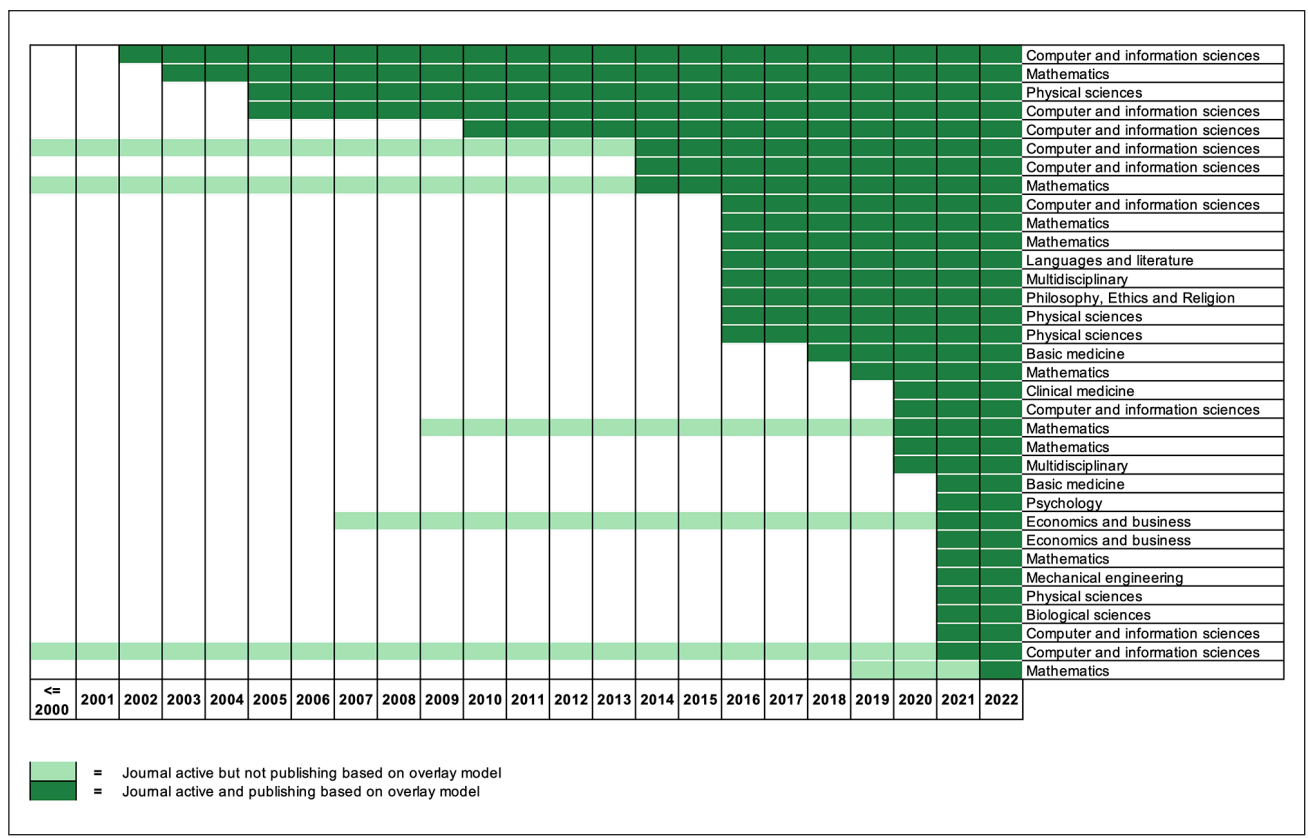

**Figure 2.** Year of conversion to overlay model between 2000 and 2022; one row per journal as indicated by research discipline. Potential period of publishing before the overlay model is depicted in lighter shade.

### RQ2.4: What are the policies for submissions and preprints?

All of the investigated overlay journals operated with the principle of open submissions, that is, none of the journals operated with only invited submissions. Twenty-five (74%) of the investigated journals only accepted submissions of preprints openly available in repositories prior to being sent to the journal. Although 9 (26%) of the overlay journals also accepted direct submissions of manuscripts that were not available in preprint repositories, these journals often encouraged authors to upload their manuscript to a repository prior to submission.

### RQ2.5: Which repositories are journals using as part of an overlay model?

arXiv and HAL were the most frequent repositories utilised by the investigated overlay journals. Twenty-six (76%) of the investigated journals allowed submissions from arXiv and 17 (50%) allowed manuscript submissions from HAL. Figure 3 presents the primary field of research of the investigated overlay journals and the distribution of repositories used for enabling the overlay model of operation.

### RQ2.6: What are the five largest overlay journals by article volume and how has their output developed over the last 4 years?

Although the investigated overlay journals most often represented computer and information sciences or mathematics, the journal with the currently largest article volume publishes on physical sciences; the *Open Journal for Quantum Science* had the largest number of articles published during 2018–2021. Figure 4 presents the top five journals with the most articles published during 2018–2021.

### RQ2.7: What are the publication languages?

All 34 journals accepted manuscripts in English. Eight of these journals also accepted manuscripts in French, two of which also allowed German and one allowed Spanish and Italian. The strong presence of French as a publication language is likely connected with Episciences, a French overlay service provider to many of the overlay journals (see also RQ 2.10). Fifteen of the overlay journals were based in France, five were based in the United States, three in Canada, two in the United Kingdom and two in Germany. Overlay journals were also based in Austria (1),

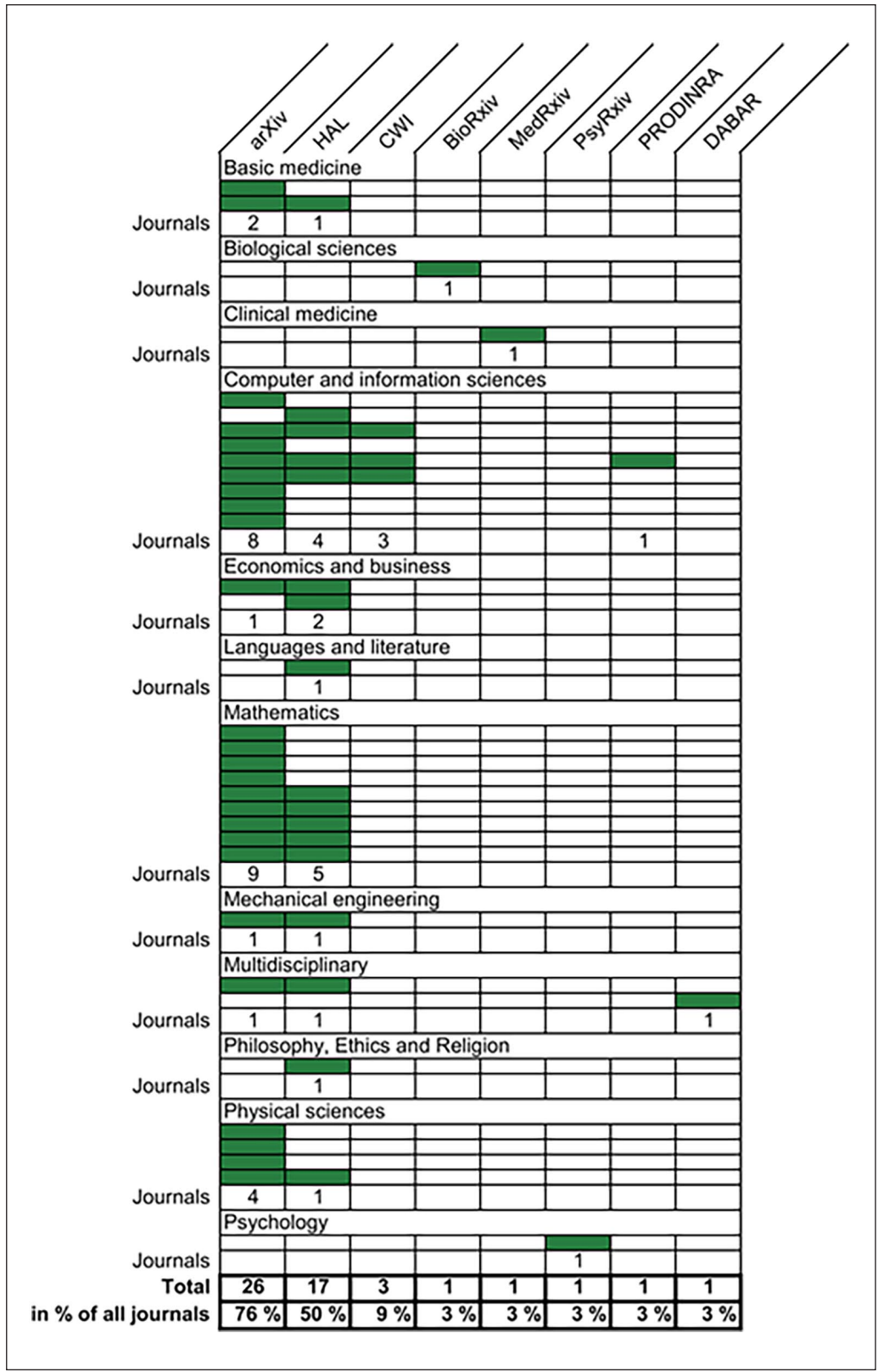

**Figure 3.** Preprint repository use of overlay journals, sorted by journal discipline categories. Each row represents one journal.

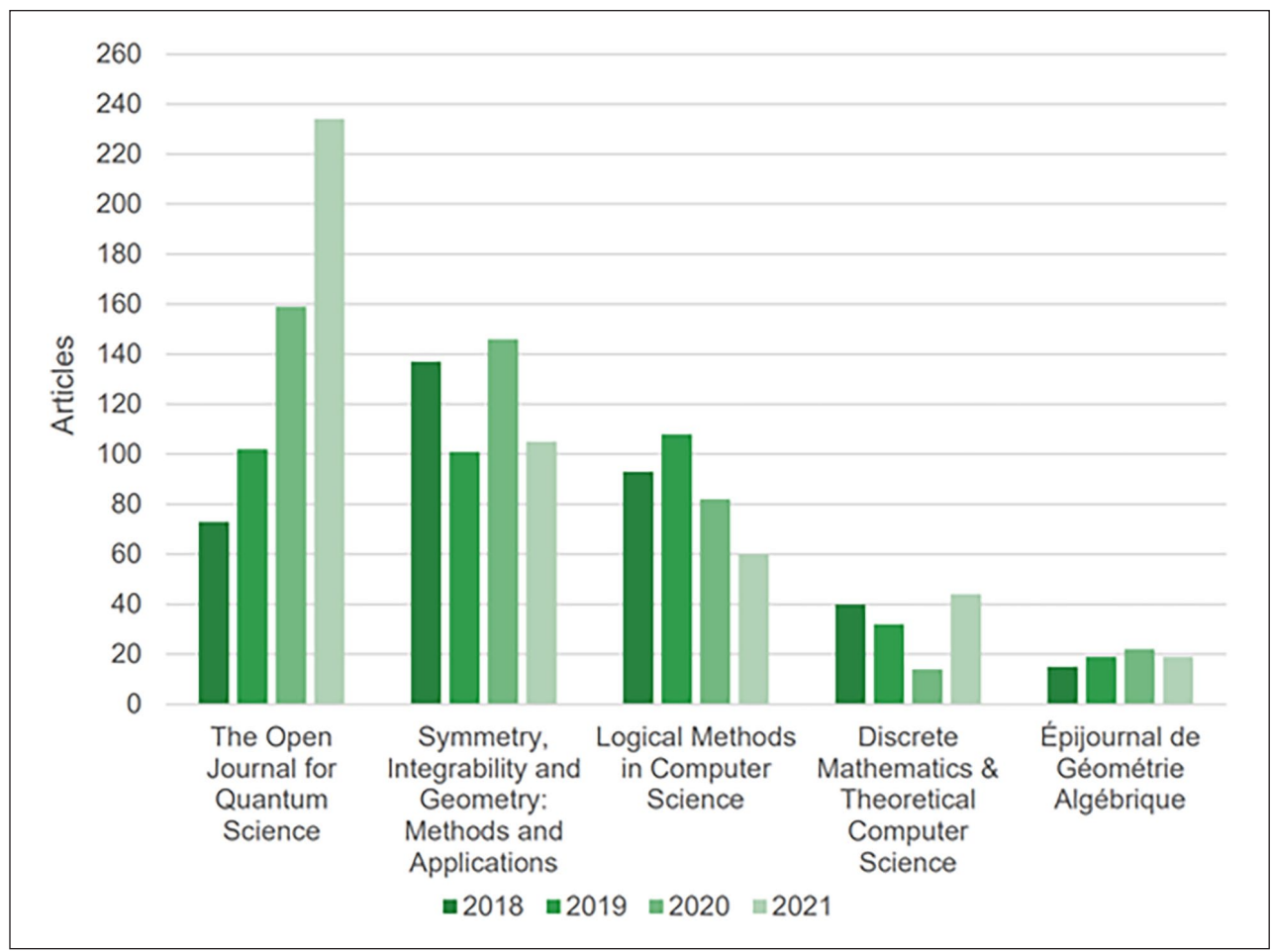

**Figure 4.** Article volume development in the top five overlay journals by volume during 2018–2021 (*Fundamenta informaticae*, which converted to an overlay model towards the end of 2021, was excluded from consideration in this analysis).

Croatia (1), India (1), Ireland (1), Netherlands (1), Romania (1) and Ukraine (1).

### RQ2.8: What are the publisher types?

The data concerning publisher types revealed that more than half ($n = 18$; 53%) of the investigated overlay journals were published by groups of academics (journals' editorial boards). In other words, the journal was published using a preprint repository as a manuscript source and managed through an overlay journal service provider (e.g. Episciences.org or Scholastica), but no formal publisher information was found. Table 2 presents the publisher types of the investigated overlay journals.

### RQ2.9: What overlay journal service providers are used?

Episciences was the most common service provider for the investigated journals. In total, 20 (59%) of the investigated journals utilised the Episciences' platform for overlay journals, while 8 (24%) used Scholastica's services.

**Table 2.** Publisher types of the investigated overlay journals ($n = 34$).

| Publisher type | Share of overlay journals |
| --- | --- |
| Individual academics or group of academics | 18 (53%) |
| Scientific society or professional association | 5 (15%) |
| Commercial | 4 (12%) |
| University, university department, research institute | 4 (12%) |
| Professional, non-commercial publisher | 2 (6%) |
| University press | 1 (3%) |

### RQ2.10: What licences are used for published content?

Most ($n = 25$; 74%) of the investigated overlay journals published their articles under Creative Commons (CC) licences (Table 3). It is noteworthy that although all of the investigated journals were OA journals, 7 (21%) did not specify the licence or copyright information, which would determine the reuse terms of the published articles.

**Table 3.** Publishing licences of the investigated overlay journals (n = 34).

| Publishing licence | Share of overlay journals |
| --- | --- |
| Creative commons CC-BY | 16 (47%) |
| Licence not defined | 7 (21%) |
| Creative commons (unspecified) | 5 (15%) |
| Creative commons CC-BY-SA | 3 (9%) |
| Creative commons CC-BY-NC-SA | 1 (3%) |
| Variant of arXiv.org supported licences with authors retaining copyright | 1 (3%) |
| arXiv.org non-exclusive licence to distribute | 1 (3%) |

### RQ2.11: How well are the journals indexed in DOAJ, Scopus and WoS?

About one-third of the investigated overlay journals (n = 10; 29%) were indexed in DOAJ, while a similar amount (n = 11; 32%) were indexed in Elsevier's Scopus database and Clarivate analytics' Web of Science database (n = 10; 29%). The low number of indexed journals in the prior services is at least partly explained by the fact that 13 (38%) of the journals were founded during 2020 or after, or had not yet issued their first volume. Although a minority (n = 11; 32%) of journals were indexed in the Scopus database, the overall number of articles from overlay journals indexed in Scopus is increasing.

### RQ2.12: What do the journal metrics look like?

The analysis revealed that the Nordic journal rankings are not consistent in their classifications of the investigated overlay journals. Thirteen (38%) of the investigated journals had a Finnish Publication Forum class for 2021, and all 13 (38%) were considered as standard or 'basic' level journals (level 1). Respectively, 11 (32%) of the investigated journals were classified as level 1 in the Norwegian Register for Scientific Journals, Series and Publisher for 2021. Notably, only six of the investigated overlay journals were classified in the 2021 Danish Bibliometric Research Indicator list of journals (four journals being classified as level 1; two journals as level 2). Appendix 1 compiles the observed journal metrics for the investigated overlay journals.

The overlay journals representing natural sciences fared the best in the Nordic journal rankings. From the born-overlay journals, only *Logical Methods in Computer Sciences* was classified as a level 2 journal in the Nordic journal rankings and received this classification in the 2021 Danish Bibliometric Research Indicator. *Fundamenta Informaticae* was also classified as level 2 in the Danish Bibliometric Research Indicator 2021 list of journals. However, as the journal only recently converted to the overlay model, its classification derives from the era utilising a different publishing model.

The analysis revealed that overlay journals may also rank highly within the traditional journal citation metrics. About 5 (15%) of the investigated journals had an Impact Factor metric for the year 2020 (as released in 2021), although the Impact Factor Rankings of *Fundamenta Informaticae* journals was derived from its era before converting to the overlay model. *The Open Journal for Quantum Science* ranked the highest within the field of science-specific Impact Factor rankings (see Appendix 1).

## Discussion and conclusions

This study was able to identify 34 overlay journals through an explorative data collection process. The findings suggest that overlay publishing is still in its early stages, hence there are few overlay journals currently available. In comparison, the DOAJ directory currently indexes more than 16,000 OA journals (DOAJ, 2021) and the Ulrichsweb database lists more than 24,000 OA journals (Ulrichsweb, 2021). However, the findings show that new overlay journals have been actively established within recent years; almost half of the investigated overlay journals (n = 16, 47%) started as overlay journals from 2020 or later. The scarcity of prior studies examining the number of overlay journals makes it difficult to determine whether the amount of overlay journals has substantially risen within, for example, the past 10 years. This study strives to remedy this situation by publishing the data in its entirety (https://doi.org/10.5281/zenodo.6420517). This study could be undertaken again after 5 years to examine how the landscape of overlay journal publishing has developed over time.

Most of the current overlay journals represent natural sciences – in particular, computer and information sciences, physics and mathematics. In addition, notwithstanding the journals that recently transformed to the overlay model, all overlay journals with Nordic quality rankings and Impact Factors came from the abovementioned fields of science. This prevalence of journals representing natural sciences is not surprising given that these fields developed the first open preprint repositories (Brown, 2001). Although the most prominent overlay journals came from natural sciences, the analysis revealed that there are emerging journals also, for example, within the fields of basic medicine, biological sciences and economics and business (see also Thornton and Kroeker, 2021). Given that Tennant et al. (2017: 21) reported in 2017 that the overlay publishing model was not adopted beyond natural sciences, this recent development where other fields have also started to experiment with the model is noteworthy.

It is possible that the observed increase in recently established overlay journals occurred in conjunction with the growth of subject repositories and popularity of preprint uploads in more research disciplines. Though it has been found that many traditional journals still have unclear policies for preprint availability of submitted manuscripts (Klebel et al., 2020), the situation has been improving over time, which also gives authors increased confidence in distributing their work before or during submission to a journal. Some journals, including the portfolio of journals from the Public Library of Science (PLOS) now even incorporate direct transfer and/or facilitated posting of manuscripts to bioRxiv and medRxiv (PLOS.org, 2022). Within recent years, open repositories have also been established in fields such as chemical engineering (chemRxiv) (see Chiarelli et al., 2019). In addition, it is noteworthy that the HAL repository, which was the second most frequent article source for the investigated overlay journals, is a multidisciplinary repository with a scope beyond natural sciences.

The current overlay journals are commonly published by groups of scientists rather than formal organisations. One of the key motivations behind establishing these journals appears to be to combat the rising costs of scientific publishing, particularly those arising from subscription fees and OA article processing charges (e.g. Räsänen, 2019; Scholastica, 2019). As suggested by the findings here, the current overlay journals seem to live up to this expectation: none of the investigated journals required payment of fees from the authors or readers, which is at least partly enabled by the cost-effective aspects of the overlay publishing model. Moreover, the cost-effectiveness of the model may encourage academics to start publishing overlay journals on their own without formal commercial or non-commercial publishers.

Basing journal operations mostly on voluntary time may also bring risks to journal operations, which are not exclusive to the overlay journals but concern OA publishing more broadly. In a comprehensive study of all journals included in the DOAJ, Crawford (2021) found that 60% of all OA journals were published by university publishers, which often operate at a small scale relative to professional publisher organisations. In a recent study of OA journals that do not collect fees from authors and readers, 60% of responding journals reported high or medium reliance on volunteers (Bosman et al., 2021). The reliance on volunteers may pose a threat for sustainability of operations if key individuals leave the journal. Laakso (2021) surveyed journal editors running Nordic OA journals, and found that people and organisational aspects were key themes in the answers from the respondents. Without predictable funding streams or a persistent backing organisation, journal operations may become vulnerable. In order to understand specific challenges faced by overlay journals, it is important to study why journals such as *BiOverlay* or *Journal of High Energy Physics* have either deactivated or transferred to another publishing model. In general, the overlay movement could benefit from sharing best editorial practices of, for example, *The Open Journal of Quantum Science*, which is able to publish a yearly volume of over 200 articles without requiring fees from either authors or readers.

A detailed analysis of the costs of individual overlay articles or the model are beyond the scope of this article. The operating costs of arXiv (Cornell University, 2022) and overlay journal web services such as Scholastica (2021) and Episciences.org (2021) suggest that the overlay model is cost efficient when compared to the traditional subscription-based publishing (Grossmann and Brembs, 2021; Schimmer et al., 2015). However, given the current diminutive amount of overlay journals, the larger scale cost-saving effects of the model seem to be linked to the willingness of researchers to adopt this model of publishing in the future.

Citation impact comparisons between OA and subscription journals is a complex issue involving discipline-specific considerations (Torres-Salinas et al., 2019: 141). As pointed out by Björk and Solomon (2012), researchers may find impactful OA journals by choosing their journals carefully. The findings of this article suggest that this principle also applies to overlay journals. The *Open Journal for Quantum Science* was ranked highly within the field of science-specific Impact Factor rankings. Furthermore, it is noteworthy that the Impact Factor rankings are derived retrospectively (i.e. the journal's Impact Factor as released in 2020 is based on the citation data of 2017–2018). As 15 (44%) of the investigated journals were founded after 2017, this excludes them from even plausibly having an Impact Factor metric released in 2020.

In a recent study of preprints available in bioRxiv, Anderson (2020) found that most manuscripts had been uploaded to the repository close to or after submission to a journal, suggesting that preprints were not dominantly used as a mechanism for pre-submission feedback. Based on the study, one-third of preprints had not been published as peer-reviewed journal articles within 2 years of upload. Considering these results in the context of the process flows that could be observed from the policies of the overlay journals included in this study, there is nothing inherent in the common implementation of overlay model publishing that would change this pattern. Submission to a repository can be done just before submitting to the journal (most journals simply ask for a link to the preprint in the submission form) or not at all if the journal also accepts direct submissions but uses repository functions for archiving published content (as was the case for 9 out of 34 journals). Combined with the low presence of open peer-review practices among these journals, this suggests that the model does not provide a radical change to the open science practices as part of the pre-publication and review process, as is commonly implemented today.

## Conclusions

The overlay journal movement is still in its early stages. It could be argued that this 'movement' is better described as a diverse set of ways through which journals integrate repository archival into their workflows – sometimes to facilitate open science practices early on in the submission stages, and other times simply for robust archival of accepted and published content. The current presence of overlay journals is diminutive compared to the overall number of OA journals, but our findings show that new overlay journals have been actively established recently. The research fields for overlay journals are skewed towards journals within computer science and informatics, mathematics and the physical sciences, however recently established overlay journals also include medical, biological and multidisciplinary research. Overlay journals are commonly published by groups of scientists rather than formal organisations, and these journals may rank highly within the traditional journal citation metrics. None of the investigated journals required payment of fees from the authors or readers, which is at least partly enabled by the cost-effective aspects of the overlay publishing model. Moreover, the cost-effective aspects of the overlay model may encourage academics to start publishing overlay journals on their own without formal commercial or non-commercial publishers. Both the adoption of OA preprint repositories and researchers' willingness to publish in overlay journals will determine the model's wider impact on scientific publishing.

## Acknowledgements

The authors thank Dr. Tommi Tenkanen for his important comments and feedback considering this work. We also thank the Scholastica and PubPub teams for kindly providing information of their overlay journals. It is important to note that the authors are solely responsible for any mistakes found from this work. The authors would like to thank the anonymous reviewers for their constructive comments that greatly helped to improve the manuscript. In addition, the authors thank Dr. Martyn Rittman for his open review of our work. Dr. Jacquelin De Faveri helped with the English editing.

## Declaration of conflicting interests

The author(s) declared the following potential conflicts of interest with respect to the research, authorship, and/or publication of this article: Open access provided by Aalto University.

## Data availability statement

The collected data are provided as open data at https://doi.org/10.5281/zenodo.6420517 to facilitate future research.

## Funding

The author(s) received no financial support for the research, authorship, and/or publication of this article.

## ORCID iDs

Antti Mikael Rousi  https://orcid.org/0000-0002-4184-7035
Mikael Laakso  https://orcid.org/0000-0003-3951-7990

## Author biographies

Antti Mikael Rousi is a senior advisor at Aalto University, Finland. He has worked extensively with open science, publishing and current research information systems. He holds a PhD in information sciences.

Mikael Laakso is an associate professor at Hanken School of Economics, Helsinki, Finland. He has been researching the changing landscape towards openness in scholarly publishing by studying combinations of bibliometrics, web metrics, business models, science policy, and author behavior for over 10 years.

**Appendix 1.** Nordic journal rankings, journal citation impact and impact factor metrics of the investigated overlay journals (*n* = 34).

| Journal | FPF class (2021) | DBRI class (2021) | NRSJP class (2021) | JCI (2020) | IF (2020) | IF rank (2020) | Research discipline | Note |
|---|---|---|---|---|---|---|---|---|
| Advances in Combinatorics | 1 | – | – | – | – | – | Mathematics | |
| Discrete Analysis | 1 | – | – | 1.12 | – | – | Mathematics | |
| Discrete Mathematics and Theoretical Computer Science | 1 | – | 1 | 0.27 | 0.596 | 104/108 Computer science, software engineering; 279/330 Mathematics | Computer and information sciences | |
| Épijournal de Géométrie Algébrique | – | – | 1 | 0.53 | – | – | Mathematics | |
| Fundamenta Informaticae | 1 | 2 | – | 0.76[a] | 1.333[a] | 78/108 Computer science, software engineering; 140/265 Mathematics, applied[a] | Computer and information sciences | Transferred to overlay model in 2021 |
| Hardy Ramanujan Journal | 1 | – | 1 | – | – | – | Mathematics | |
| Internet Mathematics | 1 | 1 | 1 | – | – | – | Mathematics | |
| Journal of Data Mining and Digital Humanities | 1 | 1 | – | – | – | – | Computer and information sciences | |
| Journal of Groups, Complexity, Cryptology | – | – | 1 | 0.33[a] | – | – | Mathematics | Transferred to overlay model in 2020 |
| Journal of Philosophical Economics | 1 | 1 | – | 0.14[a] | – | – | Economics and business | Transferred to overlay model in 2021 |
| Logical Methods in Computer Science | 1 | 2 | 1 | 0.42 | 0.438 | 105/110 Computer science, theory and methods; 15/21 Logic | Computer and information sciences | |
| Symmetry, Integrability and Geometry: Methods and Applications | 1 | 1 | 1 | 0.36 | 1.072 | 38/55 Physics, mathematical | Physical sciences | |
| The Art, Science and Engineering of Programming | 1 | – | 1 | – | – | – | Computer and information sciences | |
| The Open Journal for Quantum Science | 1 | – | – | 0.67 | 6.777 | 4/17 Quantum science and technology; 9/86 Physics, multidisciplinary | Physical sciences | |
| The Open Journal of Astrophysics | 1 | – | 1 | – | – | – | Physical sciences | |
| Total amount of journals with metrics (% from all journals) | 13 (38%) | 6 (18%) | 11 (32%) | 9 (26%)[a] | 5 (15%)[a] | 5 (15%)[a] | Computer and information sciences (*n* = 5); Economics and business (*n* = 1); Mathematics (*n* = 6); Physical sciences (*n* = 3) | |

[a]The journals of *Fundamenta Informaticae*, *Journal of Groups, Complexity, Cryptology* and *Journal of Philosophical Economics* were recently converted to the overlay model, and their Journal Citation Impact, Impact Factors and Impact Factor Rankings are derived from their era before converting to the overlay model.